\documentclass[preprint,superscriptaddress,amsmath,amssymb]{revtex4-2}

\usepackage{graphicx}
\usepackage[utf8]{inputenc}
\usepackage{color}
\usepackage{xcolor}

\usepackage{commath} 
\usepackage{natmove}   
\usepackage{hyperref}

\usepackage{setspace}  

\setcitestyle{super}

\makeatletter
\renewcommand{\paragraph}{%
  \@startsection{paragraph}{4}%
  {\z@}{1.5ex \@plus 1ex \@minus .2ex}{-1em}%
  {\normalfont\normalsize\bfseries}%
}
\makeatother

\begin{document}

\title{Why soft contacts are stickier when breaking than when making them}

\author{Antoine Sanner}
\affiliation{%
Department of Microsystems Engineering (IMTEK), University of Freiburg, Georges-Köhler-Allee 103, 79110 Freiburg, Germany
}%
\affiliation{%
Cluster of Excellence livMatS, Freiburg Center for Interactive Materials and Bioinspired Technologies, University of Freiburg, Georges-Köhler-Allee 105, 79110 Freiburg, Germany
}
\affiliation{%
Institute for Building Materials, ETH Zurich, Zurich, 8093, Switzerland
}%

\author{Nityanshu Kumar}
\affiliation{School of Polymer Science and Polymer Engineering, The University of Akron, Akron, Ohio 44325, USA}
\affiliation{Science \& Technology Division, Corning Incorporated, Corning, NY 14831, USA}
\author{Ali Dhinojwala}
\affiliation{School of Polymer Science and Polymer Engineering, The University of Akron, Akron, Ohio 44325, USA}

\author{Tevis D. B. Jacobs}
\affiliation{Department of Mechanical Engineering and Materials Science, University of Pittsburgh, 3700 O’Hara Street, Pittsburgh, Pennsylvania 15261, USA}

\author{Lars Pastewka}
\email{lars.pastewka@imtek.uni-freiburg.de}
\affiliation{%
Department of Microsystems Engineering (IMTEK), University of Freiburg, Georges-Köhler-Allee 103, 79110 Freiburg, Germany
}%
\affiliation{%
Cluster of Excellence livMatS, Freiburg Center for Interactive Materials and Bioinspired Technologies, University of Freiburg, Georges-Köhler-Allee 105, 79110 Freiburg, Germany
}

\maketitle

{\bf
Insects, pick-and-place manufacturing, engineered adhesives, and soft robots employ soft materials to stick to surfaces even in the presence of roughness.
%
Experiments show that the force required for making contact is lower than for releasing it, a phenomenon known as the adhesion hysteresis.~\cite{chaudhury_direct_1991,dalvi_linking_2019}
The common explanation for this hysteresis is either contact aging or viscoelasticity.~\cite{israelachvili_intermolecular_1991,chen_molecular_1991}
%
Here, we show that adhesion hysteresis emerges even for perfectly elastic contacts and in the absence of contact aging and viscoelasticity because of surface roughness.
We present a crack-perturbation model~\cite{gao_nearly_1987,sanner_crack-front_2022,wei_weight_1989} and experimental observations that reveal discrete jumps of the contact perimeter.
These stick-slip instabilities are triggered by local differences in fracture energy between roughness peaks and valleys.
Pinning of the contact perimeter~\cite{larkin_pinning_1979,robbins_contact_1987,demery_microstructural_2014} retards both its advancement when coming into contact and its retraction when pulling away.
Our model quantitatively reproduces the hysteresis observed in experiments and allows us to derive analytical predictions for its magnitude,
accounting for realistic rough geometries across orders of magnitude in length scale~\cite{gujrati_combining_2018,gujrati_comprehensive_2021}.
Our results explain why adhesion hysteresis is ubiquitous and reveal why soft pads in nature and engineering are efficient in adhering even to surfaces with significant roughness.
}

\begin{figure}
    \centering
    \includegraphics{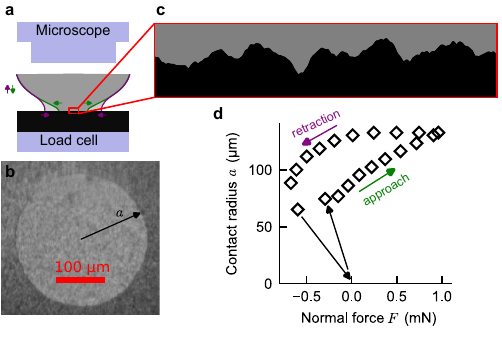}
    \caption{{\bf Phenomenology of adhesive contact.}
      \textbf{(a)}
      Many contacts can be described as spheres making contact with a flat surface.
      For soft materials, microscopic interactions are strong enough that the solids deform significantly near the contact edge.
      \textbf{(b)}
      The contact forms a circle for contacting spheres, and its radius $a$ can be measured from in-situ optical images of the contact area.
      \textbf{(c)} Most natural and technical surfaces are rough so that the solid needs to elastically deform to come into conforming contact.
      \textbf{(d)}~%
      The contact radius is larger and the normal force is more adhesive (negative) during retraction than during approach.
%
The pull-off force is the most negative force on these curves.
}
    \label{fig:concept}
\end{figure}

\paragraph*{Introduction}
Two solids stick to each other because of attractive van-der-Waals or capillary interactions at small scales.~\cite{israelachvili_intermolecular_1991}
The strength of these interactions is commonly described by the intrinsic work of adhesion $w_\text{int}$, the energy that is gained by these interactions per surface area of intimate contact.
The work of adhesion is most commonly measured from the pull-off force $F_\text{pulloff} = -3\pi w_\mathrm{int} R/2$ of a soft spherical probe (see Fig.~\ref{fig:concept}a) with radius $R$ which makes a circular contact with radius $a$ (see Fig.~\ref{fig:concept}b).~\cite{johnson_surface_1971}.
For hard substrates, the measured \emph{apparent} work of adhesion is smaller than  the \emph{intrinsic} value $w_\mathrm{int}$ because roughness limits the area of intimate contact to the highest protrusions~\cite{pastewka_contact_2014,persson_nature_2004}.
Soft solids are sticky because they can deform to come into contact over a large portion of the rough topography.
The overall strength of the adhesive joint is then determined by the balance of the energy gained by making contact and the elastic energy spent for conforming to the surface.
Following 
Persson and Tosatti~\cite{persson_effect_2001}, energy conservation implies that surface roughness reduces the apparent work of adhesion to 
\begin{equation}
    w_\mathrm{PT}=w_\text{int} - e_\text{el},
    \label{eq:persson-tosatti}
\end{equation}
where $e_\text{el}$ is the elastic energy per unit contact area required to conform to the roughness (Fig.~\ref{fig:concept}c).
As shown in Fig.~\ref{fig:concept}d, experiments typically follow different paths during approach and retraction, leading to different apparent different values for work of adhesion of adhesion for making and breaking contact, $w_\mathrm{appr}$ and $w_\mathrm{retr}$.
This \emph{adhesion hysteresis} contradicts Persson and Tosatti's balance of energy, which gives the same value $w_\mathrm{PT}$ for approach and retraction.

In this letter, we present and experimentally validate a theory that allows us to predict these apparent work of adhesion during approach and retraction and thereby the adhesive hysteresis.
For soft spherical probes, we can describe the circular contact perimeter as a crack.
The crack front is in equilibrium when Griffith's criterion is fulfilled:~\cite{griffith_vi._1921}
The energy per unit area required locally for opening the crack, $w_\mathrm{loc}$, is equal to the energy released from the elastic deformation,
$G \delta A=w_\mathrm{loc} \delta A$,
where $\delta A$ is the contact area swept out by the moving crack front.
A more common way of writing this equation is
\begin{equation}
    G=w_\mathrm{loc},
    \label{eq:force-balance}
\end{equation}
where both the energy release rate $G$ and $w_\mathrm{loc}$ should be interpreted as forces per unit crack length.
Johnson, Kendall and Roberts (JKR)~\cite{johnson_surface_1971} derived the expression for the energy release rate $G$ for a smooth spherical indenter, $G=G_\text{JKR}(b,a)$.
Equation~\eqref{eq:force-balance} then allows the evaluation of not just the pull-off force, but of all functional dependencies between rigid body displacement $b$, contact radius $a$ and normal force $F$ during contact. 

For smooth spheres, $w_\mathrm{loc}$ is the intrinsic work of adhesion $w_\mathrm{int}$ which is uniform along the surface.
In the presence of roughness, we will show below that $w_\mathrm{loc}$ becomes a spatially fluctuating field describing the topographic roughness.
Equation~\eqref{eq:force-balance} must then hold independently for each point on the contact perimeter.

\begin{figure*}[htbp]
 \includegraphics[width=\textwidth]{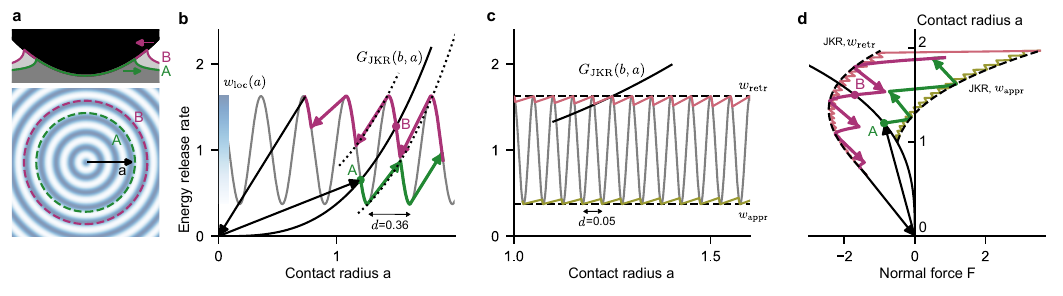}
 \caption{
  {\bf Simplified axisymmetric contact demonstrating the physical origin of adhesion hysteresis.
    }
    The indenter is a perfect sphere with axisymmetric heterogeneity in local adhesion $w_\mathrm{loc}(a)$.
\textbf{(a)}
  Cross-section of the contact at rigid body penetration $b = 0$ (top) and top view of the axisymmetric work of adhesion heterogeneity $w_\mathrm{loc}(a)$ (bottom).
  The blue color indicates regions of high adhesion.
 \textbf{(b)}
  Elastic energy release rates in an approach-retraction cycle for a sinusoidal work of adhesion  $w_\mathrm{loc}(a)$ with wavelength $d=0.36$ (gray line).
  The black line shows the elastic energy release rate $G_\mathrm{JKR}(b,a)$ as a function of contact radius for fixed rigid body penetration $b=0$.
  Fluctuations of $w_\mathrm{loc}(a)$
  lead to several metastable states A, B at fixed $b$.
  During approach, the contact perimeter is pinned in metastable states with low adhesion (green curve),
  while
  during retraction the contact perimeter is pinned at higher radii by adhesion peaks (red curve).
  Arrows indicate elastic instabilities where the contact radius jumps between metastable states.
  \textbf{(c)} Energy release rates in an approach-retraction cycle for a work of adhesion heterogeneity with smaller wavelength $d=0.05$.
  Note that the slope of $G_\mathrm{JKR}$ appears to be flatter than in panel (b) because we show a smaller range of contact radii.
  For short wavelengths, the works of adhesion sampled during approach (light green curve) and retraction (light red curve)
  stay close to the constant values $w_\mathrm{appr}$ and $w_\mathrm{retr}$.
  \textbf{(d)} The contact radius and the normal force during an approach-retraction cycle for wavelength $d=0.36$ (darker colors) and $d=0.05$ (lighter colors).
  The dashed lines are the prediction by the JKR theory using $w_\mathrm{retr}$ and $w_\mathrm{appr}$ for the work of adhesion.
  The solid black line corresponds to increasing energy release rates at fixed rigid body penetration $b=0$.
  Energy release rates are displayed in units of the average work of adhesion and lengths and forces have been nondimensionalized following the conventions of Refs.~\cite{barthel_adhesive_2008,maugis_contact_2010} as described in the Supplemental Material.
}
  \label{fig:1d-pinning}
\end{figure*}
\paragraph*{Axissymmetric chemical heterogeneity}
We first demonstrate the physical origin of the adhesion hysteresis using a simplified surface that has concentric rings of high and low adhesion energy, similar to the model by Guduru~\cite{guduru_detachment_2007-theory} and Kesari and Lew~\cite{kesari_role_2010,kesari_effective_2011}. 
Rather than being random, $w_\mathrm{loc}(a)$ varies in concentric rings of wavelength $d$ as a function of distance $a$ from the apex of the contacting sphere (Fig.~\ref{fig:1d-pinning}a).
Figure~\ref{fig:1d-pinning}b shows $w_\mathrm{loc}(a)$ alongside $G_\text{JKR}(b,a)$ for a fixed displacement $b$.
Because of the spatial variations of $w_\mathrm{loc}$, there are multiple solutions to Eq.~\eqref{eq:force-balance} indicated by the labels A and B.
Moving into contact from the solution denoted by A leads to an instability where the solution A disappears, at which the contact radius jumps to the next ring of $w_\mathrm{loc}(a)$.
This samples the lower values of $w_\mathrm{loc}$ shown by the green line in Fig.~\ref{fig:1d-pinning}b.
Conversely, moving out of contact progresses along a different path that samples the higher values of $w_\mathrm{loc}(a)$, shown by the red line.
The combination of fluctuations in $w_\mathrm{loc}$ and the elastic restoring force $G_\text{JKR}$ acts like a ratchet resisting the growing and shrinking of the contact area and leads to a stick-slip motion of the contact line.
The line is pinned by the first strong-enough obstacle it encounters, so that it is pinned at low contact radius when the contact area grows and at high radius when it shrinks.

In the limit of roughness with small wavelength, $d\to 0$, 
$G_\text{JKR}$ does not decrease significantly before the contact line arrests at the next peak (see Fig.~\ref{fig:1d-pinning}c).
In this limit, the contact line samples the minimum values $w_\text{appr}$ of $w_\mathrm{loc}$ during approach and the maximum values $w_\text{retr}$ during retraction.
The functional relationship between $b$, $a$ and $F$ then becomes identical to the JKR solution for smooth bodies (Suppl. Eqs.~(S-4)~to~(S-7)), but with a work of adhesion that is decreased during approach ($w_\text{appr}$) and increased during retraction ($w_\text{retr}$, see Fig.~\ref{fig:1d-pinning}d).
In this limit, the hysteresis $w_\text{retr} - w_\text{appr}$ becomes equal to the peak-to-peak amplitude of $w_\mathrm{loc}(a)$.~\cite{kesari_role_2010}

\begin{figure*}
    \centering
    \includegraphics[width=\textwidth]{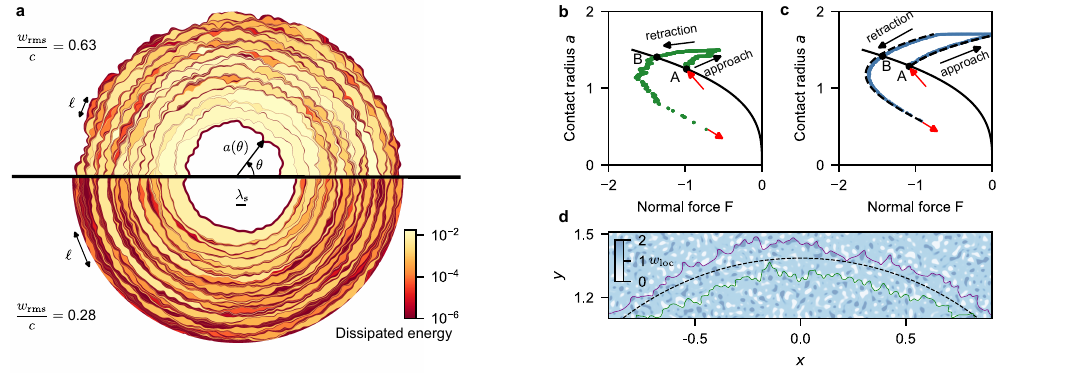}
    \caption{{\bf 
        Crack-front pinning by two-dimensional random heterogeneity.
      }
\textbf{(a)}
Evolution of the contact line during retraction in a crack-front simulation on 2D random work of adhesion field.
Each colored patch corresponds to an elastic instability during which the perimeter jumps between two pinned configurations (dark lines),
and the color scale represents the energy dissipated during each instability.
The larkin length $\ell$ corresponds to the smallest extent of these jumps along the perimeter, and increases for
weaker heterogeneity or for a stiffer line.
The work of adhesion heterogeneity corresponds to a random roughness that has a flat power spectral density with short-wavelength cutoff $\lambda_{s}=0.07$.
\textbf{(b)}~Contact radius as a function of the normal force in the simulation for the stronger pinning field shown in panel (a).
The elastic instabilities correspond to sudden jumps in the contact area and in the normal force.
The solid black line corresponds to increasing energy release rates at fixed rigid body penetration b = 0,
and the points A and B show that the contact radius is higher during retraction
than during approach.
The red arrows show the jump-in and jump-out of contact instabilities.
\textbf{(c)}~Contact radius as a function of the normal force in a simulation on a random chemical heterogeneity with smaller feature size $\approx 0.01$ and $w_\mathrm{rms} / c \approx 0.45$.
The force-radius curve is smooth because the random field has small features
that trigger a large number of instabilities.
 The dashed lines are JKR curves with work of adhesion $w_\mathrm{appr}$ and $w_\mathrm{retr}$ predicted by our theory Eq.~\eqref{eq:hysteresis_roughness}.
In this simulation, $w_\mathrm{loc}$ corresponds to a self-affine randomly rough topography with an elastic energy for fully conformal contact $e_\mathrm{el} / w_\mathrm{int} = 0.05$
and power spectrum shown by the blue circles in the inset of Suppl. Fig.~S-5.
 \textbf{(d)}
 Contact lines at rigid body penetration $b=0$ on the random work of adhesion heterogeneity shown by the blue colormap.
Floppy lines are pinned at higher contact radii during retraction (purple line) than during retraction (green line)
because they meander predominantly between regions of low adhesion (white patches) during approach,
and between regions of high adhesion (dark blue patches) during retraction.
In the limit of a rigid line, the perimeter is perfectly circular (dashed line),
randomly sampling as many regions of low and high adhesion.
%
%
Units are nondimensionalized following the conventions of Refs.~\cite{barthel_adhesive_2008,maugis_contact_2010} as described in the Supplemental Material.
    }
    \label{fig:2d-pinning}
\end{figure*}

\paragraph*{Random chemical heterogeneity}
The next step in complexity is moving from a simplified axisymmetric surface to a surface with random variation of the local work of adhesion, where the contact line is no longer perfectly circular (see Fig.~\ref{fig:2d-pinning}a).
The energy release rate $G$ at a given point now depends on the whole shape of the contact $a(s)$, where $s$ is the length of the corresponding path along the contact circle.
Based on the crack-perturbation theory by Gao and Rice~\cite{rice_first-order_1985,wei_weight_1989,gao_nearly_1987}, we recently derived the approximate expression~\cite{gao_nearly_1987,sanner_crack-front_2022}
\begin{equation}
    G(s) = G_\mathrm{JKR} (a(s))
    +
    c
    (-\Delta_s)^{1/2} a(s),
    \label{eq:G_first_order_mt}
\end{equation}
where the fractional Laplacian $(-\Delta_s)^{1/2}$ of $a(s)$ penalizes excursions from circularity  and can be interpreted as a generalized curvature, similar to the restoring force of an elastic line.
Supplementary Section S-ID derives this expression and shows that near equilibrium, where $G(s)=w_\mathrm{int}$, the stiffness of the line is given by $c = w_\text{int}$. Note that counterintuitively, the stiffness of the line does not depend on the elastic modulus of the bulk. 

Numerical solution of Eqs.~\eqref{eq:force-balance} and \eqref{eq:G_first_order_mt} (see Supplementary Section~S-II)
on a random field $w_\mathrm{loc}(x,y)$ with lateral correlation of length $d$
yields force-area curves similar to those of our axisymmetric model (Fig.~\ref{fig:2d-pinning}b,c).
The key difference is that the contact line now advances and recedes in jumps  (Fig.~\ref{fig:2d-pinning}a) that are localized over a characteristic length $\ell$, the Larkin length.~\cite{larkin_pinning_1979,imry_random-field_1975,robbins_contact_1987,monti_sliding_2020,demery_microstructural_2014}
Between these jumps, the contact line is pinned.
At the same rigid body penetration, pinning occurs at lower contact radii in approach than during retraction, leading to a hysteresis in apparent adhesion described by two JKR curves with constant apparent work of adhesion $w_\text{appr}$ and $w_\text{retr}$ (Fig.~\ref{fig:2d-pinning}c), similar to the curves obtained from our 1D axisymmetric model (Fig.~\ref{fig:1d-pinning}d).

Our numerical data in Suppl. Fig.~S-5 shows that the magnitude of hysteresis, $w_\text{retr} - w_\text{appr} \propto w_\mathrm{rms}^{2}$, the variance of the random field $w_\mathrm{loc}$.
To understand this expression, we first discuss the virtual limit $c\to 0$ where the line is floppy and deviations from circularity are not penalized.
 Floppy lines ($c < w_\mathrm{rms}$)
 can freely distort and meander along valleys during approach (green line in Fig.~\ref{fig:2d-pinning}d) and peaks during retraction (purple line).
Because of this biased sampling of the work of adhesion along the line,
the contact radius is larger during retraction than during approach.
In this individual-pinning limit~\cite{joanny_model_1984,patinet_quantitative_2013,demery_microstructural_2014}, each angle $\theta$ along the contact perimeter independently yields our 1D model and we obtain $w_\mathrm{retr} - w_\mathrm{appr}\propto w_\mathrm{rms}$.
In the opposite limit, $c\to\infty$ the line is stiff and the contact remains circular (dashed line),
randomly sampling as many regions of low and high adhesion.
The fluctuations average out along the perimeter so that there is no hysteresis,
$w_\mathrm{retr} - w_\mathrm{appr} = 0$.
The contact radius is obtained from the JKR expression evaluated for the spatially averaged work of adhesion, $\langle w_\mathrm{loc} \rangle$.

Our simulations (and experiments as shown below) are in an intermediate regime, characterized by local jumps over length $\ell$ or $N=\ell/d$ pinning sites.
The line is stiff over the Larkin length $\ell$ and hence samples a coarse-grained work of adhesion field $w^{(\ell)}$ with $w^{(\ell)}_\text{rms} = w_\mathrm{rms}/\sqrt{N}$.
From force-balance Eq.~\eqref{eq:G_first_order_mt} we obtain that an excursion of the contact line by distance $\delta a$ over this length leads to a restoring force $\delta G \propto c \delta a/\ell$, which must balance $w^{(\ell)}_\text{rms}$.
We note that $\delta a \approx d$, which is the distance to the closest local stable configuration.~\cite{larkin_pinning_1979,robbins_contact_1987}
The equilibrium condition $\delta G= w_\mathrm{rms}^{(\ell)}$ then yields
\begin{equation}
  \label{eq:larkin-n}
    N \propto \left(c/w_\mathrm{rms}\right)^2
\end{equation}
or $\ell = Nd$.
This means, the magnitude of the hysteresis must scale as
\begin{equation}
    w_\mathrm{retr} - w_\mathrm{appr} \propto w_\mathrm{rms}^{(\ell)} \propto w_\mathrm{rms}^2/c,  
    \label{eq:whyst}
\end{equation}
exactly as observed in our simulations.
Identical results were obtained previously for cracks in heterogeneous media.~\cite{demery_effect_2014,demery_microstructural_2014}

\begin{figure}
    \centering
    \includegraphics{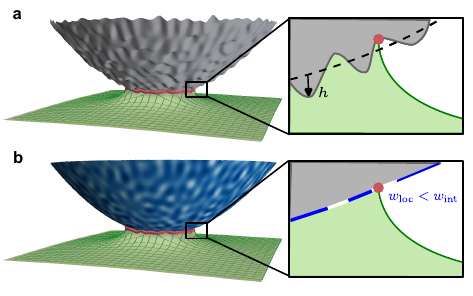}
    \caption{
    {\bf Mapping topographic roughness to equivalent chemical heterogeneity.}
    The contact of a rough sphere (a) is equivalent to the contact of a sphere with a work of adhesion heterogeneity $w_\mathrm{loc}$ (b).
    The solid is stretched at the crack tip and surface roughness perturbs this elastic deformation.
    The associated perturbation of the elastic energy 
    can equivalently be described by fluctuations of the work of adhesion.
    }
    \label{fig:contact-edge}
\end{figure}

\paragraph*{Topographic roughness} The final step in describing adhesion hysteresis on real surfaces is to relate the randomly rough topography to the spatial variations in local work of adhesion. 
For this we need to consider excursions of the contact line normal to the surface in additional to the lateral excursions that are described by the contact radius $a(\theta)$ (see Fig.~\ref{fig:contact-edge}).
First note that the solid is always dilated near the crack tip.
In order to conform to a valley, the elastic solid needs to stretch even more, requiring elastic energy.
Using the same arguments that lead to Eq.~\eqref{eq:persson-tosatti}, this additional elastic energy manifests as an effectively decreased \emph{local} work of adhesion.
Conversely, conforming to a peak decreases the overall strain near the crack tip and releases elastic energy, leading to an increased equivalent work of adhesion.
While this intuitive picture approximately describes the relationship between heights and local adhesion,
the quantitative value of the local adhesion $w_\mathrm{loc}$ depends nonlocally
on the topographic field $h(x, y)$ via an integral transformation  derived in Suppl. Sec. I B and C.
Supplementary Section III also shows that a crack-front simulation on $w_\mathrm{loc}(x,y)$ yields results virtually indistinguishable from an exact boundary-element calculation.

\paragraph*{Comparison to experiments}

\begin{figure*}
 \includegraphics[width=\textwidth]{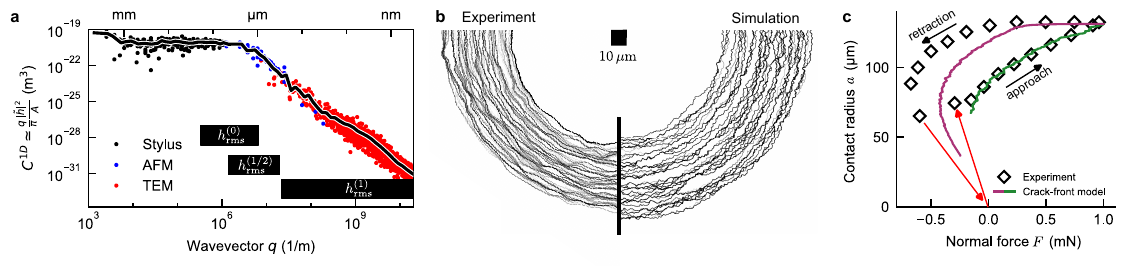}
 \caption{{\bf Crack-front pinning explains adhesion hysteresis on real-world surfaces with multiscale roughness.}
   \textbf{(a)}
  Power spectral density (PSD) of a nanocrystalline diamond (NCD) film extracted from more than 60 measurements of Ref.~\cite{gujrati_comprehensive_2021},
  combining stylus profilometry, atomic force microscopy (AFM), and transmission electron microscopy (TEM).
  Black bars indicate the range of scales that dominates $h^{(\alpha)}_\mathrm{rms}$ (Eq.~\eqref{eq:rms-derivatives}).
  Specifically, short wavelengths and long wavelengths beyond this range respectively contribute to only 10\% of the value of $(h^{(\alpha)}_\mathrm{rms})^2$.
  Evaluating Eq.~\eqref{eq:rms-derivatives} requires the 2D or isotropic power-spectral density of the surface topography, while only the 1D PSD is known.
  Following Refs.~\cite{jacobs_quantitative_2017,dalvi_linking_2019},
  we converted the 1D PSD $C^\mathrm{1D}$ to the isotropic 2D PSD using the approximation $C^\mathrm{iso}(q) \simeq \frac{\pi}{q} C^\mathrm{1D}(q)$.
  The data used in this figure is available online in Ref.~\cite{nanocrystalline_diamond_v2}.
  \textbf{(b)}
  Position of the perimeter in the contact between a rubber sphere and a rough surface during approach.
  The perimeters on the left side are extracted from the experiment on NCD shown in Fig.~\ref{fig:concept},  and the right side shows equilibrium positions of the perimeter in a crack perturbation simulation (see Supplementary material I and II)  on random roughness similar to NCD.
  The contact perimeter is pinned where the black lines are close to each other,
  while regions with low density of lines indicates where the contact perimeter accelerates during an instability.
  The simulation predicts instabilities of various sizes, reaching a lateral extent up to several tens of $\mathrm{\mu m}$.
In the experiment, only the largest instabilities and the largest features of the contact line are visible because of the limited resolution of the camera and because we removed image noise using a spatially averaging filter. 
  Details on the extraction of the position of the line from the video are described in the Methods section and the original video recording is provided in the Supplementary Video SV1.
  The positions of the perimeter are shown from jump into contact until the force reaches 0.64$\,$mN.
\textbf{(c)}
   Contact radius and normal force during approach and retraction of the experiment (diamonds) and simulation (continuous line) shown in panel (b).
    We extracted the intrinsic work of adhesion $w_\text{int}=63\,\text{mJ/m}^{2}$ used in the simulation by fitting the work of approach. 
The sphere has Young's modulus $E=0.7\,\text{MPa}$
and radius $R=1.25\,\text{mm}$.
More details on the experimental setup are provided in Methods and in Ref.~\cite{dalvi_linking_2019}.
Supplementary Material Fig.~S-6 shows that the power-spectral density of the synthetic random roughness used in the simulation is close to the power-spectral density of NCD at the length scales that dominate $h^{(1/2)}_\mathrm{rms}$.
  }
  \label{fig:roughness}
\end{figure*}

We contacted a rough nanodiamond film with a PDMS hemispherical lense while optically tracing the contact perimeter (see Methods).
The nanodiamond film was characterized from atomic to macroscopic length scales using a variety of techniques, as described in Refs.~\cite{gujrati_combining_2018,gujrati_comprehensive_2021}.
The resulting power-spectral density (PSD)~\cite{jacobs_quantitative_2017} comprehensively describes the topography of the film and is shown in Fig.~\ref{fig:roughness}a.
This experiment is compared to a simulation carried out on roughness field with an identical PSD, leaving $w_\text{int}$ as the only free parameter that we fit to the approach curve.
This yields $w_\text{int} = 63~\text{mJ}~\text{m}^{-2}$, within the range expected for van-der-Waals interaction. 

First, our experiments show the same instabilities as the simulations.
The trace of the contact line in Fig.~\ref{fig:roughness}b shows the jerky motion of the line for both, with comparable amplitudes of deviations from the ideal contact circle.
Videos of the contact area in the indentation experiment show stick-slip motion of the contact line, similar to our simulations (Suppl. Mat. SV1).
The fundamental hysteresis mechanism in our model, elastic instabilities and stick-slip motion of the contact line, is clearly present in the experiment.

Second, measurements of the mean contact radius as a function of normal force also agree with our simulation results (Fig.~\ref{fig:roughness}c).
While the intrinsic work of adhesion was adjusted such that the simulations follow the experimental data during approach, the overall functional form is JKR-like (with an effective $w_\text{appr}$) and agrees between experiments and simulation.
During retraction, we observe the same phenomenology: From the point of largest normal force, the sphere retracts first at constant contact radius before starting to follow a JKR-like curve with an increased work of adhesion $w_\text{retr}$.
While simulations retract at slightly different forces, the order of magnitude of the hysteresis is correctly predicted from our simple elastic model.

Quantitative differences could come from intrinsic assumptions in our model, such as small strains, linear elastic properties, approximations made in deriving the simplified crack-front expressions, or the assumption of conforming contact.
Increasing roughness increases adhesion only as long as the energy needed to fully conform the surface roughness $e_\text{el}$ is lower than the gain in surface energy $w_\text{int}$~\cite{persson_nature_2004,mulakaluri_adhesion_2011,wang_is_2022}.
Many experiments that report a decrease of pull-off force with increasing roughness, as for example reported in the classic adhesion experiment by Fuller and Tabor~\cite{fuller_effect_1975}, may be in this limit where only partial contact is established within the contact circle.
Unlike the theory presented here for soft solids and our understanding of nonadhesive contact~\cite{persson_nature_2004}, there is presently no unifying theory that quantitatively describes adhesive contact for stiff solids.
Large scale simulations with boundary-element methods are needed to better understand this intermediate regime.~\cite{medina_numerical_2014,popov_strength_2017,pastewka_contact_2014,wang_is_2022}

\paragraph*{Analytic estimates}
We now show that simple analytic estimates can be obtained from our crack-front model.
The equivalent work of adhesion field has the property that its mean corresponds to the Persson-Tosatti expression,  Eq.~\eqref{eq:persson-tosatti}.
Furthermore, it has local fluctuations with amplitude $w_\mathrm{rms} =  \sqrt{2 w_\text{int} e_\text{el}}$ which determine the adhesion hysteresis, which means that the main parameter determining the hysteresis is $e_\text{el}$, see Eq.~\eqref{eq:whyst}.
We carried out crack-front simulations on self-affine randomly rough topographies (Fig.~3c and Suppl.~Sec.~IV) with varying parameters to confirm that the work of adhesion during approach and retraction is indeed given by
\begin{equation}
\label{eq:hysteresis_roughness}
w_\mathrm{\overset{\scriptstyle retr}{ \scriptstyle appr}}
=
w_\mathrm{int} - e_\mathrm{el} \pm k e_\mathrm{el}
,
\end{equation}
with a numerical factor of $k\sim 3$.

%
The elastic energy for fully conformal contact can be written as
\begin{equation}
\label{eq:eel}
e_\text{el} = \frac{E^\prime}{4}\left[h^{(1/2)}_\text{rms}\right]^{2},
\end{equation}
where $E^\prime$ is the elastic contact modulus~\cite{johnson_contact_1985} and $h^{(1/2)}_\text{rms}$ is a geometric descriptor of the rough topography.
In terms of the two-dimensional PSD~\cite{jacobs_quantitative_2017} $C^\mathrm{iso}$, we define
\begin{equation}
\label{eq:rms-derivatives}
    (h_\text{rms}^{(\alpha)})^2
    =
    \frac{1}{4\pi^2}
    \int \dif^2 q \, |\vec{q}|^{2\alpha} C^{\mathrm{iso}}(|\vec{q}|)
    ,
\end{equation}
where $\vec{q}$ is the wavevector.
This expression contains the root-mean-square (rms) amplitude of the topography, $h_\text{rms}^{(0)}$, the rms gradient of the topography, $h_\text{rms}^{(1)}$, as well as arbitrary derivatives of order $\alpha$.
The elastic energy is given by the roughness parameter $h^{(1/2)}_\mathrm{rms}$, which is intermediate between rms heights and rms gradients.

For most natural and engineered surfaces, $h^{(1/2)}_\mathrm{rms}$ depends on the large scales,
like the rms height, because their Hurst exponent $H>0.5$~\cite{mandelbrot_fractal_1984,candela_roughness_2012,persson_effect_2001,persson_fractal_2014}.
Our model is then consistent with the increase in pull-off force with $h_\text{rms}$ reported in Refs.~\cite{briggs_effect_1976,kesari_role_2010}.
We note that most measurements report insufficient details on surface roughness to allow definite conclusions on the applicability of a certain contact model.
The range of length scales that dominate $h^{(1/2)}_\text{rms}$ in our own experiments is at the transition between power-law scaling and the flat rolloff at $2$~$\mu$m, a length scale that is accessible with an atomic-force microscope.
We illustrate the respective scales that contribute to $h_\text{rms}^{(\alpha)}$ in Fig.~\ref{fig:roughness}a.

\paragraph*{Summary \& Conclusion}

The work performed by a soft indenter during the approach-retraction cycle is dissipated in
elastic instabilities triggered by surface roughness.
%
%
%
%
The dissipated energy is the difference in energy between the pinned configurations just before and just after the instability (see Fig.~\ref{fig:2d-pinning}a).
This pinning of the contact line
explains why adhesion is always stronger when breaking a soft contact than when making it,
even in the absence of material specific dissipation. 
Roughness peaks increase local adhesion,
which pins the contact line and increases the pull-off force.
By describing rough adhesion as the pinning of an elastic line,
we were able to derive parameter-free, quantitative expressions for the hysteresis in terms of a simple statistical roughness parameter.
This analysis paves the way to better understanding the role of surface roughness in adhesion, and provides guidance for which scales of roughness to control in order to tune adhesion.

\section*{Methods}

\paragraph*{Rough substrate}

We contacted the PDMS lens against a nanocrystalline diamond (NCD) film of known roughness.
The diamond film was deposited on a silicon wafer by chemical vapor deposition and subsequently hydrogen terminated to avoid polar interactions and hydrogen bond formation between the PDMS lens and the rough substrate.
The roughness of the film was determined by combining measurements from the milimeter to atomic scales using stylus profilometer, atomic force microscopy (AFM), and transmission electron microscopy (TEM).
The full experimental dataset along with the averaged PSD shown in Fig.~\ref{fig:roughness}b are available online~\cite{nanocrystalline_diamond_v2}.
Details on the film growth and the multiscale topography characterization are provided in Refs.~\cite{gujrati_combining_2018,gujrati_comprehensive_2021}.
\paragraph*{Synthesis of PDMS hemispheres}

We synthesized PDMS hemispheres of 0.7$\,$MPa Young's modulus by hydrosilylation addition reaction.
Vinyl-terminated PDMS V-41 (weight-averaged Molar mass $M_w = 62,700\,\text{g/mol}$) as monomer, tetrakis-dimethylsiloxysilane as tetra-functional cross-linker and platinum carbonyl cyclo-vinyl methyl siloxane as catalyst were procured from Gelest Inc. Monomer and cross-linker were first mixed in a molar ratio of 4.4 in an aluminum pan.
The catalyst was added as 0.1 weight percent of the total mixture, and finally the batch was degassed in a vacuum chamber for 5 minutes.
Hemispherical lenses were cast on fluorinated glass dishes using a needle and a syringe,
and cured at 60° C for three days.
Since the PDMS mixture has a higher surface energy than the fluorinated surface, the drops maintain a contact angle on the surface, giving a shape of a hemispherical lens. 
We extracted the radius of curvature $R=1.25\,$mm from a profile image of the lens.
After curing reaction, the lenses were transferred to cellulose extraction thimble for Soxhlet extraction where toluene refluxes at 130° C for 48 hrs. PDMS lenses were again transferred to a fluorinated dish and dried in air for 12 h. Finally, the lenses were vacuum dried at 60° C for 16 h and then used for experiments. The radius of curvature was measured by fitting a 3-point circle to the image obtained using an optical microscope (Olympus).

We determined Young's modulus $E=0.7~$MPa by fiting the JKR theory to an indentation experiment against a flat silicon wafer covered with octadecyltrichlorosilane (OTS).
This experiment also shows that in the absence of surface roughness, the work of adhesion hysteresis is below $10\,$mJ/m$^{2}$, a value significantly smaller than the hysteresis measured against NCD ($\simeq 80\,$mJ/m$^{2}$).

\paragraph*{Indentation experiment}

We measured the force and area during approach and retraction of a PDMS hemisphere against a rough diamond film using the setup of Dalvi et al.~\cite{dalvi_linking_2019}.
The lens and the substrate were approached at a constant rate of 60 nm/s until a repulsive force of 1\,mN and then retracted with the same rate.
The PDMS hemisphere is transparent, allowing simultaneous measurement of the force and of the contact area, Fig.~\ref{fig:concept}b.
The video recording, provided in the Suppl. Mat. SV1, has a frame interval of 0.3s, but Fig.~\ref{fig:roughness}c shows values for the force and contact radius at intervals of $\approx 30\,$s.

\paragraph*{Extraction of contact line from video}

We extracted the perimeter from each time frame of the video of the contact area.
The contact area appears as a bright region in the video, and we defined the contact perimeter as a contour line of fixed level of gray.
At the length scale of a few pixels, the position of the line is affected by noise on the image.
To reduce the effect of noise on the position of the line,
we subtracted the image of the contact area at maximum penetration and subsequently applied a spatial Gaussian filter of variance 2 pixels.
The lines shown in Fig.~\ref{fig:roughness}b therefore only reflect the position of the perimeter on coarse scales.
Supplementary Material video SV2 shows that these lines match the shape of the contact area at large scales and follows the same intermittent motion.
The original video is available in the Suppl. Mat. SV1.

\section*{Acknowledgements}
We thank
W. Beck Andrews, 
Patrick Dondl, 
Lucas Fr\'erot, Mathias Lebihain and Mark Robbins
for insightful discussions.
The work by AS and LP was funded by the Deutsche Forschungsgemeinschaft (grant  EXC-2193/1 – 390951807) and the European Research Council (StG 757343). AD acknowledges funding by the National Science Foundation (DMR-2208464) and TDBJ acknowledges support from the National Institute for Occupational Safety and Health (R21 OH012126).



\newpage

\setcounter{figure}{0}
\setcounter{section}{0}
\setcounter{table}{0}
\setcounter{equation}{0}

\renewcommand{\thesection}{S-\Roman{section}}
\renewcommand{\thefigure}{S-\arabic{figure}}
\renewcommand{\thetable}{S-\arabic{table}}
\renewcommand{\theequation}{S-\arabic{equation}}

\renewcommand{\L}[1] {}
\newcommand{\M}[1] {}
\newcommand{\MP}[1]{}
\newcommand{\pargoal}[2]{}

\newcommand{\aeq}[0]{{ {a_\text{eq}}}}

\begin{center}
  {\Large\bf{ Supplementary Material for \\
   ``Why soft contacts are stickier when breaking than when making them'' }}
  
  \vspace{1cm}
  
  {\bf Antoine Sanner$^{1,2,3}$,  Nityanshu Kumar$^{4, 5}$, Ali Dhinojwala$^{4}$, Tevis D.B. Jacobs$^{6}$
    and Lars Pastewka$^{1,2}$}
  \end{center}
  
  $^1$ Department of Microsystems Engineering (IMTEK), University of Freiburg, Georges-Köhler-Allee 103, 79110 Freiburg, Germany
  
  $^{2}$ Cluster of Excellence \emph{liv}MatS, Freiburg Center for Interactive Materials and Bioinspired Technologies, University of Freiburg, Georges-Köhler-Allee 105, 79110 Freiburg, Germany
  
  $^{3}$ Institute for Building Materials, ETH Zurich, Zurich, 8093, Switzerland
  
  $^4$ School of Polymer Science and Polymer Engineering, The University of Akron, Akron, Ohio 44325, USA
  
  $^{5}$ Science \& Technology Division, Corning Incorporated, Corning, NY 14831, USA
  
  $^6$ Department of Mechanical Engineering and Materials Science, University of Pittsburgh, 3700 O’Hara Street, Pittsburgh, Pennsylvania 15261, USA

  \section{Crack-front model}
  
\begin{figure}
    \centering
  \includegraphics[width=\textwidth]{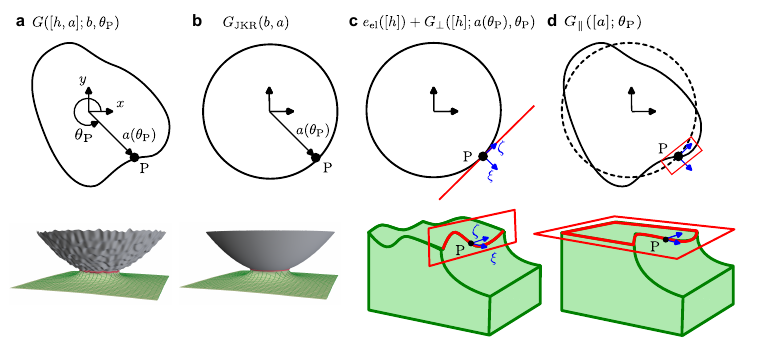}
  \caption{
    We consider the contact of a sphere of radius $R$ with roughness $h(x,y)$ superposed to it.
    {\bf (a)} Because of surface roughness, the contact perimeter is no longer circular.
     We describe it by the contact radius $a(\theta)$, the planar distance between the tip of the sphere and the perimeter of the contact.
     The energy release rate $G$ at the point P along the crack is decomposed into three contributions, $G=G_\text{JKR} + (e_\text{el} + G_\perp) + G_\parallel$, illustrated in panels (b) to (d).
     {\bf (b)} The energy release rate $G_\text{JKR}(b,a)$ for the smooth contact is given by the theory of Johnson, Kendall and Roberts~\protect\cite{johnson_surface_1971}.
    {\bf (c)} Surface roughness leads to out-of-plane displacements of the contact perimeter. This increases the average energy release rate by $e_\text{el}$ and leads to additional local fluctuations $G_{\perp}([h];a(\theta_{\mathrm{P}}),\theta)$. Here, $e_\text{el}$ is the elastic energy needed to fully conform to surface roughness.
    {\bf (d)} The in-plane deflection of the perimeter from circularity leads to the additional contribution $G_{\parallel}([a],\theta)$.
  }
    \label{fig:concept-perp-parallel}
  \end{figure}

  Our goal is to model the contact of a rough sphere on a deformable elastic flat (Fig.~\ref{fig:concept-perp-parallel}a).
  The contact perimeter of the adhesive contact between a smooth sphere and flat can be regarded as a circular crack (Fig.~\ref{fig:concept-perp-parallel}b).
  This is the basis of the Johnson, Kendall and Roberts (JKR) model for adhesion~\cite{johnson_surface_1971}.
  JKR derived an expression for the elastic energy release rate $G_\text{JKR}$ for this spherical geometry, and balanced it with the intrinsic work of adhesion, $G_\text{JKR}=w_\text{int}$.
  Here, we extend this result to rough spheres, where the crack shape deviates from circularity.
  Surface roughness perturbs the shape of the crack in the surface normal direction (Figs.~4 and \ref{fig:concept-perp-parallel}c).
  This perturbs the local balance of energy, leading to additional deviation of the crack shape in direction parallel to the surface (Fig.~\ref{fig:concept-perp-parallel}d).
  
  Figure~\ref{fig:concept-perp-parallel} illustrates this decomposition in terms of the energy release rate $G$.
  The surface roughness $h$ locally perturbs the elastic energy by $ G_{\perp}$ (Fig.~\ref{fig:concept-perp-parallel}c) \cite{anderson_stress_1987} and the perimeter distorts within the plane to satisfy equilibrium with the uniform work of adhesion $w_\text{int}$ (Fig.~\ref{fig:concept-perp-parallel}d).
  As we show in detail in this supplementary material, we describe the effect of surface roughness by an equivalent work of adhesion field
  \begin{equation}
  \label{eq:W-equivalent}
  w_\mathrm{loc}([h]; a(\theta), \theta) = w_\text{int} - e_\text{el}([h]) - G_{\perp}([h]; a(\theta), \theta),
  \end{equation}
  where $e_\text{el}$ is the elastic energy required to fully conform to the surface roughness and the square brackets indicate a functional dependency.
  
  The effect of the in-plane deflection on the elastic energy  $G_{\parallel}$ was derived by Gao and Rice~\cite{gao_nearly_1987} and later extended by us to spheres \cite{sanner_crack-front_2022}.
  In our simulations, the equilibrium condition
  \begin{equation}
  \label{eq:equilibrium-W-G}
  w_\mathrm{loc}([h]; a(\theta), \theta) = G_\text{JKR}(b, a(\theta)) + G_{\parallel}([a], \theta) 
  \end{equation}
  determines the contact radius with $\mathcal{O}(h^{2})$ errors in the strength of the disorder.
  The left hand side represents the driving force to increase the contact radius that fluctuates according to the surface roughness, while the right hand side represents the elastic response of the line that only depends on the spherical geometry and the material properties.
  The numerical implementation follows Refs.~\cite{rosso_roughness_2002,sanner_crack-front_2022} and is summarized in supplementary material~\ref{sec:numerical-cf}.
  We validate our equations by comparing crack-front simulations to boundary element method simulations in supplementary material~\ref{sec:cf-bem}.
  Equations~\eqref{eq:W-equivalent}~and~\eqref{eq:equilibrium-W-G}
  establish the equivalence between the adhesion of rough spheres and the classic problem of the pinning of an elastic-line by quenched disorder~\cite{imry_random-field_1975,
  larkin_pinning_1979,
  robbins_contact_1987,
  rosso_roughness_2002,
  demery_microstructural_2014,
  demery_effect_2014}.
   
  \subsection{Axisymmetric contact: The JKR model}
  \label{sec:JKR}
  
  We consider the contact of a sphere (to be exact, a paraboloid) adhering an elastic half-space at a fixed rigid body penetration $b$ (Fig.~\ref{fig:concept-perp-parallel}a).
  This case can be mapped to the contact of two spheres with the same composite radius $R$
  and contact modulus $E^\prime$ \cite{johnson_contact_1985}.
  When only one half-space deforms, $E^\prime=E/(1-\nu^2)$, where $E$ is Young's modulus and $\nu$ is Poisson's ratio.
  Fracture mechanics typically considers the contact of two elastic half-spaces where $E^\prime=E / 2 (1-\nu^2)$.
  We assume the contact is frictionless and consider only vertical displacements of the half space.

  The equilibrium radius and force for a perfect sphere against the axisymmetric work of adhesion heterogeneity $w_\mathrm{loc}(a)$ is given by the JKR theory~\cite{johnson_surface_1971,barthel_adhesive_2008,maugis_contact_2010}.
  JKR described the adhesion of a paraboloid with radius $R$ by superposing the displacement and the stress fields of the nonadhesive Hertzian contact~\cite{hertz_ueber_1881} and the circular flat punch under tensile load~\cite{sneddon_boussinesqs_1946}.
  
  The contact pressures $p$ have a tensile singularity as the distance to the edge of the contact $-\xi$ goes to 0,
  \begin{equation}
  \label{eq:10}
  p(\xi) = - K_\mathrm{JKR}/\sqrt{2\pi (-\xi)} + \mathcal{O}((-\xi)^{1/2})
  ,
  \end{equation}
  with the stress intensity factor
  \begin{equation}
  \label{eq:4}
  K_\text{JKR} = \left( \frac{a^2}{R} - b\right) \frac{{E^\prime}}{\sqrt{\pi a}}
  .
  \end{equation}
  {Here and below we use the subscript JKR to indicate the circular contact to a smooth sphere.}
  The energy release rate depends solely on the amplitude of this singularity \cite{irwin_analysis_1957}
  \begin{equation}\label{eq:Irwin}
    G_\mathrm{JKR} = K_\mathrm{JKR}^{2}/(2 E^\prime)
  \end{equation}
  and the equilibrium condition $G_\text{JKR}(b, a) = w_\mathrm{loc}(a)$ yields the contact radius.
  The normal force is given by
  \begin{equation}\label{eq:jkr_force}
  F_\mathrm{{JKR}}(a, b) = \frac{4 E^\prime}{3 R} a^3 + 2aE^\prime \left(b - \frac{a^2 }{R}\right).
  \end{equation}
  At equilibrium, where $G_\mathrm{JKR}$ = $w_\mathrm{loc}$, the relationship between force and contact area is given by
  \begin{equation}\label{eq:jkr_force_equilibrium}
    F_\mathrm{{JKR}}(a) = \frac{4 E^\prime}{3 R} a^3 - \sqrt{8 \pi w_\mathrm{loc}(a) E^\prime a ^{3}}
    .
  \end{equation}
  
  Once nondimensionalized using distinct vertical and lateral length units, the JKR contact is parameter free \cite{muller_influence_1980, maugis_contact_2010, muser_single-asperity_2014}, 
  and we present our numerical results in the nondimensional units defined in Refs.~\cite{barthel_adhesive_2008,maugis_contact_2010}.
  %
  Specifically,
  lengths along the surface of the half-space (e.g., the contact radius) are normalized by $(3\pi w_\text{int} R^2 / 4 E^\prime )^{1/3}$,
  lengths in vertical direction (e.g., displacements) by $(9\pi^2 w_\text{int}^2 R / 16 {E^\prime}^2)^{1/3}$
  and normal forces by $\pi w_\text{int} R$.
  The equations are in dimensional form but can be nondimensionalized by substituting $R=1$, $w_\text{int} = 1/\pi$ and $E^\prime = 3 / 4$.

  \subsection{Circular contact with surface roughness: Out-of-plane perturbation of the elastic energy release rate}
  \label{sec:effect-of-roughness}
  
  We now determine the energy release rate at the perimeter of the contact with a \emph{rough} sphere but where the contact perimeter remains circular~(Fig.\ref{fig:concept-perp-parallel}c).
  We denote the respective energy release by $G_{\circ}([h]; b, a, \theta)$, where the brackets indicate a functional dependency on the height field $h(x,y)$ that describes the roughness.
  Out-of-plane deflections of the surface of the solid make the elastic energy release rate $G_{\circ}([h]; b, a, \theta)$ fluctuate along the contact perimeter, and $\theta$ parameterizes the angle along the perimeter of the circular crack front.
  In the main text and in our simulations,
  we formally describe this perturbation of the energy release rate by the equivalent work of adhesion $w_\mathrm{loc}$.
  In order to justify this mapping, we first discuss the \emph{true} elastic energy release rate $G_{\circ}$ and show that the effects of the spherical geometry, surface roughness and in-plane distortion of the crack-front are decoupled.

  The JKR contact is the superposition of the (adhesive) flat punch~\cite{sneddon_boussinesqs_1946}
  and the Hertz solution~\cite{hertz_ueber_1881}.
  For the rough sphere, we now additionally superpose the stresses and displacements needed to conform to the surface roughness.
  We do not need to determine the whole distribution of contact stresses
  because the energy release rate only depends on the stress intensity factor at the contact edge via Irwin's relation~\cite{irwin_analysis_1957}
  \begin{equation}
  \label{eq:G-Kr-KJKR}
    G_{\circ}([h];b, a, \theta)
    = \left\{K_{\text{JKR}}(b, a)+ K_\perp([h];a, \theta)\right\}^{2}/(2 E^{\prime}),
  \end{equation}
  where $K_\perp$ captures the effect of roughness.
  $K_\perp$ can be thought of as the stress intensity factor in the conforming contact of a circular flat punch with roughness $h$ at zero external load.
  Note that the stress intensity factors of the JKR solution and the influence of roughness can be superposed linearly, because in linear elasticity we can simply superpose stresses originating from different geometric contributions.
  
  \subsubsection{Stress intensity factor caused by roughness at the tip of a semi-infinite crack}
  
  \begin{figure}  
    \includegraphics[width=0.8\textwidth]{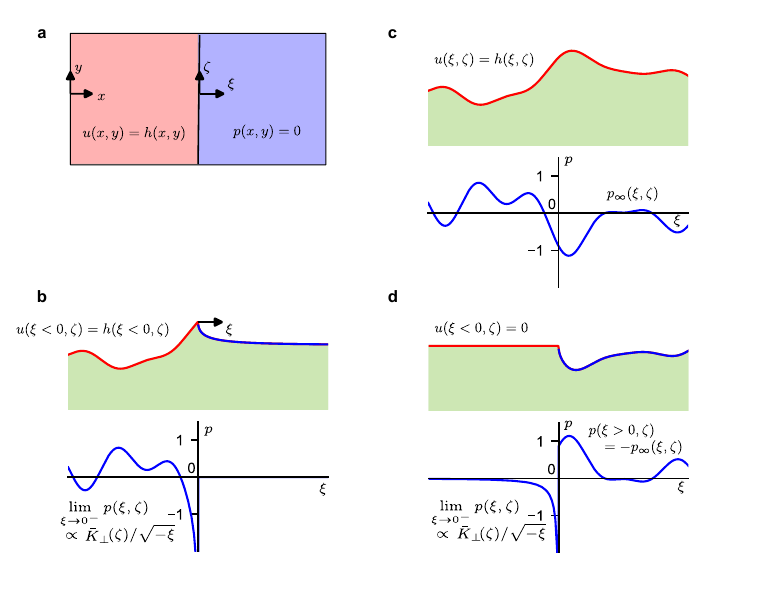}
    \caption{
    We compute the stress intensity factor 
    caused by surface roughness for a straight crack $\bar K_\perp$
    using a superposition.
    \textbf{(a)} We consider a semi-infinite crack located at $x$, for which the positive $x$ and $\xi$ directions point towards the cracked area.
    The local coordinate system $\xi, \zeta$ is centered on the crack tip, so that $\xi<0$ corresponds to the contact area.
    \textbf{(b)} 
    For $\xi > 0$, the surface is free to move vertically and the pressure $p=0$.
    For $\xi < 0$, the solid fully conforms to the surface roughness so that the displacements $u$ are prescribed to be equal to the heights $h$.
    Note that the positive direction for displacements and heights, corresponding to roughness peaks, points into the elastic halfspace (downwards).
  In the contact area, surface roughness causes fluctuating contact pressures $p(x, y)$ with stress intensity factor $\bar K_\perp(x, y)$. 
    We compute $\bar K_\perp(x, y)$ by superposing the solutions of two elastic problems (c) and (d). 
    \textbf{(c)} Displacements and pressures in an uncracked contact with the roughness $h$. 
    The displacements $u(x, y)=h(x,y)$ cause the pressure distribution $p_\infty(x, y)$
    \textbf{(d)} Semi-infinite crack with pressures applied on his crack faces.
    We apply the pressures $-p_\infty$ so that the pressures cancel out on the crack faces when superposing to (c).
    The displacements are 0 in the contact area so that the contact condition $u = h$ remains satisfied for $\xi < 0$ after superposition.
    Loading the crack faces at fixed displacements in the contact area causes the stress intensity factor.
    This stress intensity factor corresponds to $\bar K_\perp$ because there is no stress singularity in solution (c).
    }
    \label{fig:KR-superposition}
  \end{figure}
  
  We compute $K_\perp$ approximately by treating the contact as a semi-infinite crack (Fig.~\ref{fig:KR-superposition}a,b), i.e. the roughness features are small compared to the contact radius.
  %
  %
  We describe the semi-infinite crack in the coordinate system $\xi, \zeta$, where $\xi$ points in the normal to the crack front with $\xi<0$ in the contacting area.
  $\zeta$ points parallel to it.
  This is essentially a locally rotated coordinated system at the angle $\theta$ on the crack, as shown in Fig.~\ref{fig:concept-perp-parallel}c.
  The semi-infinite crack hence represents a small subsection of the circular perimeter centered at $\xi=0$ and $\zeta = 0$.

  We compute the stress intensity factor by a classic superposition \cite[2.6.4 Full Stress Field for Mode-I Crack in an Infinite Plate]{zehnder_fracture_2012}, where we first compute the pressures needed to conform the surface roughness in the absence of a crack (Fig.~\ref{fig:KR-superposition}c) and subsequently cancel out these pressures on the crack faces ($\xi>0$) (Fig.~\ref{fig:KR-superposition}d).
  Loading the crack-faces while keeping the displacements fixed in the contact area ($\xi<0$) leads to the stress intensity factor $\bar{K}_\perp(\zeta)$.
  The bar over $\bar{K}_\perp$ indicates that the result is valid for a straight crack.
  
  The pressures needed to conform to the surface roughness in the infinite contact are~\cite{westergaard_general_1935, persson_effect_2001},
  \begin{equation}
  \label{eq:p-from-u}
  \tilde{p}_\infty(\vec q) = \frac{E^{\prime}}{2} |\vec{q}| \tilde h(\vec q),
  \end{equation}
  where  $\vec q = (q_x, q_y)$ is the wavevector and the tilde denotes the Fourier transform,
  \begin{equation}
  \label{eq:3}
  \tilde h(q_x,q_y) = \int \limits_{-\infty}^{\infty} \dif x \dif y \, e^{-i (q_x x + q_y y)} h(x, y)
  .
  \end{equation}
  The stress intensity factor at the edge of the contact results from the crack-face loading needed to cancel \(p_\infty\) outside the contact area,
  \begin{equation}
  \label{eq:KR-from-pR}
  \bar{K}_\perp([h];x,y) = \int_{x}^{\infty} \dif x_\text{P} \int_{-\infty}^{\infty} \dif y_\text{P}\, k(x_\text{P}-x, y_\text{P}-y) \left\{- p_\infty([h];x_{\mathrm{P}}, y_{\mathrm{P}})\right\}
  .
  \end{equation}
  The quantity $\bar{K}_\perp$ is the stress intensity factor at position $y$ along the tip of a crack advanced to position $x$ (Fig.~\ref{fig:KR-superposition}a).
  The crack-face weight function \cite{tada_stress_2000}
  %
  \begin{equation}
  \label{eq:seminfite-CFWF}
  k(\xi, \zeta) = \frac{\sqrt{2/ \pi^3} \sqrt{\xi}}{\xi^2 + \zeta^2},
  \end{equation}
  is the stress intensity factor at the origin of a semi-infinite crack caused by a unit point force at $(\xi, \zeta)$.
  Evaluating the convolution Eq.~\eqref{eq:KR-from-pR} for each position of the crack $x$ yields a two-dimensional field of stress intensity factors, which can be most easily represented in terms of its Fourier modes,
  \begin{equation}
  \label{eq:KR-from-h-Fourier}
  \tilde{\bar{K}}_\perp(q_x, q_y) = - \frac{E^\prime}{\sqrt{2}} \sqrt{i q_x + |q_y|}  \tilde{h}(q_x, q_y)
  \end{equation}
  with
  \begin{equation}
  \label{eq:KR-Fourier-inverse-transform}
  \bar{K}_\perp([h];x, y) = \frac{1}{4\pi^{2}}
  \int \limits_{-\infty}^{\infty} \dif q_x\dif q_y\, \tilde{\bar{K}}_\perp(q_x, q_y) e^{i (q_{x} x + q_{y} y)}
  .
  \end{equation}
  Note that $\bar{K}_\perp$ has zero average (because of symmetry of the elastic surface response) and that the solids overlap where the stress intensity factor is negative.
  Our final result has no overlap provided that $|\bar{K}_\perp| < K_{\rm JKR}$.
  Anderson and Rice \cite{anderson_stress_1987} derived an equation equivalent to Eq.~\eqref{eq:KR-from-h-Fourier} to understand the interaction of crack tips with dislocations.

  We now detail the steps leading from Eq.~\eqref{eq:KR-from-pR} to Eq.~\eqref{eq:KR-from-h-Fourier}.
  The pressures needed to conform to the surface roughness in the infinite contact $p_\infty$ are easier to express in Fourier space, see Eq.~\eqref{eq:p-from-u}.
  Using the Heaviside step function $\Theta(\xi)$, we now define the weight function on the whole plane as
  \begin{equation}
  \label{eq:2}
      f(\xi, \zeta) = \Theta(\xi) k(\xi, \zeta).
  \end{equation}
  This allows us to extend the integration bound on Eq.~\eqref{eq:KR-from-pR} to infinity.
  The convolution theorem then yields the simple expression
  \begin{equation}
      \label{eq:K-p-f-Fourier}
      \tilde{\bar{K}}_\perp(q_{x}, q_{y}) = \tilde{f}^*(q_{x}, q_{y}) (-\tilde{p}_\infty(q_{x}, q_{y})),
    \end{equation}
  where the star is the complex conjugate.
  We now compute the Fourier transform of this generalized weight-function
  \begin{equation}
  \label{eq:6}
  \tilde{f}(q_{x}, q_{y}) = \int \limits_{-\infty}^{\infty} \dif \xi \dif \zeta \, \Theta(\xi) k(\xi, \zeta) e^{-iq_{x} \xi} e^{-iq_{y} \zeta}.
  \end{equation}
  Using that
  \begin{equation}
  \label{eq:7}
  \int\limits_{-\infty}^\infty \dif y\, e^{-iq_{y}y} k(x, y) = \int\limits_{-\infty}^\infty \dif y \, e^{-iq_{y}y} \frac{\sqrt{2/ \pi^3} \sqrt{x}}{x^2 + y^2} =  \sqrt{\frac{2}{\pi}} \frac{e^{- x |q_{y}|}}{\sqrt{x}}
  \end{equation}
  %
  and evaluating the step function, we get a classic Laplace transform~\cite[Eq.~29.3.4]{abramowitz_handbook_1964}
  \begin{equation}
  \tilde{f}^*(q_{x}, q_{y}) = \sqrt{2} \int \limits_{0}^{\infty} \dif x \, \frac{1}{\sqrt{\pi x}} e^{-(|q_{y}|-i q_{x}) x} = \frac{\sqrt{2}}{\sqrt{|q_{y}| - i q_{x}}}
  .
  \end{equation}
  Inserting Eq.~\eqref{eq:p-from-u} into Eq.~\eqref{eq:K-p-f-Fourier} and using that \(|q| = \sqrt{|q_{y}| - i q_{x}} \sqrt{|q_{y}|+ i q_{x}}\) yields Eq.~\eqref{eq:KR-from-h-Fourier}.
  
  \subsection{Circular contact with surface roughness: Equivalence between topographic roughness and chemical heterogeneity}
  
  
  We now switch from the straight crack back to the contact of a sphere.
  The first step is to approximate the stress intensity factor $K_\perp$ by the result for the straight crack, $\bar{K}_\perp$, obtained above.
  This approximation requires us to rotate the straight crack to be tangential to the contact circle, i.e. to rotate it by the angle $\theta$ that gives the circumferential position (Fig.~\ref{fig:concept-perp-parallel}c).
  Note that for isotropic random fields, this rotation becomes inconsequential and we do not carry it out for the results shown in the main text.
  We do carry out this rotation when comparing the crack-front results to the boundary element method, shown in Sec.~\ref{sec:cf-bem} below.
  Approximating $K_{\perp}$ by $\bar{K}_{\perp}$ is not strictly necessary,
  since
  integral expressions for the stress intensity factor for a circular flat punch with surface roughness
  exist~\cite{borodachev_contact_1991,fabrikant_stress_1998,argatov_comparison_2022}.
  The advantage of the expression for the straight crack Eq.~\eqref{eq:KR-from-h-Fourier} is that it links the statistics of the chemical heterogeneity to the statistics of the surface roughness in a simple way,
  while the expressions for the circular contact are more difficult to evaluate and to interpret.

  We now rewrite Eq.~\eqref{eq:G-Kr-KJKR} as
  \begin{equation}
      \label{eq:G-Kr-KJKR2}
      G_{\circ}([h];b, a, \theta)
      =
      G_\text{JKR}(b, a) +K_\text{JKR}(b, a) K_\perp([h];a, \theta) / E^{\prime} + K_\perp^{2}([h];a, \theta) / (2 E^{\prime}).
  \end{equation}
  The term $K_\perp$ is stochastic, as it describes the influence of surface roughness, which is typically a random field.
  Since $K_\perp$ is linear in $h$, its spatial average $\langle K_\perp\rangle_{a,\theta}$ vanishes.
  For a random field with a short correlation length, even partial averages over just the angle $\theta$ must vanish.
  This means the middle summand in Eq.~\eqref{eq:G-Kr-KJKR2} does not contribute to the average energy release rate.
  However, the variance $\langle K_\perp^2 \rangle_{a,\theta}$ must be positive and nonzero.
  Parseval's theorem tells us that,
  \begin{equation}
      \label{eq:KR-variance}
      \left\langle K^2_\perp/ 2 E^{\prime} \right\rangle_{a, \theta}
      = \frac{E^{\prime}}{16 \pi^{2}}  \int \dif q_{x} \dif q_{y} \, |q| C^\mathrm{2D}(q_x, q_y)
      = e_\text{el},
  \end{equation}
  where $C^\mathrm{2D}(q_x, q_y) = (L_x L_y)^{-1}|\tilde{h}(q_{x}, q_{y})|^{2}$ is the power spectral density of the heights~\cite{jacobs_quantitative_2017} and $L_x, L_y$ are the period of the system in the respective direction.
  %
  Note that while we consider the limit of an infinite system size $L_x, L_y \to \infty$, $C^\mathrm{2D}$ remains finite.
  %
  The variance gives the elastic energy $e_\text{el}$ for fully conformal contact.
  The average of Eq.~\eqref{eq:G-Kr-KJKR2} then becomes
  \begin{equation}
      \label{eq:average}
      \langle G_{\circ} \rangle_{a,\theta}
      =
      G_\text{JKR}
      +
      e_\text{el}.
  \end{equation}
  This equation is equivalent to a classic result by Persson and Tosatti~\cite{persson_effect_2001}.
  They approximated equilibrium by $\langle G_{\circ} \rangle_{a,\theta} = w_\text{int}$.
  Formally, this can be described by the equilibrium of a smooth sphere with the uniform equivalent work of adhesion $w_\mathrm{loc} = w_\text{int} - e_\text{el}$, where $G_\circ = G_\mathrm{JKR}$.
  This approximation only works in the adiabatic limit.
  Fluctuations become crucial when they are able to pin the crack front and trigger instabilities.
  
  We now show how to generalize Persson and Tosatti's result to describe local fluctuations.
  This means we need to consider the effect of the second term in Eq.~\eqref{eq:G-Kr-KJKR2},
  \begin{equation}
      G_{\perp}([h];b, a, \theta)
      =
      K_\text{JKR}(b, a) K_\perp([h];a, \theta) / E^{\prime},
  \end{equation}
  that disappears in the average but represents the leading-order effect of roughness on the fluctuations of $G_\circ$.
  $G_{\perp}$ depends on the geometry and position of the indenter via $K_{\mathrm{JKR}}$.
  This coupling between macroscopic boundary conditions and the microscopic disorder is a second-order effect of the roughness, which we can neglect because our final equilibrium equation determines the crack shape with first-order accuracy only.
  To first order in $h$, we approximate $K_\text{JKR}\approx\sqrt{2 w_\text{int} E'}$, yielding
  \begin{equation}
      \label{eq:G-perp-first-order}
      G_{\perp}([h];a, \theta)
      \approx  \sqrt{2 w_\text{int}/ E^{\prime}} K_\perp([h];a, \theta).
  \end{equation}
  This first-order approximation allows us to describe the effect of surface roughness by the equivalent quenched disorder in work of adhesion
  \begin{equation}
  \label{eq:W-equivalent2}
  w_\mathrm{loc}([h];a,\theta) = w_\text{int} - e_\text{el}([h])  - G_{\perp}([h]; a, \theta).
  \end{equation}
  The equivalent work of adhesion Eq.~\eqref{eq:W-equivalent2} contains only the essential leading-order contributions of the roughness
  and is independent of the macroscopic geometry,
  so that our results generalizes to other adhesion setups where our approximations are valid.

  Our mapping to an equivalent work of adhesion establishes a link to the pinning of elastic lines by quenched disorder.
  Theoretical work on the pinning of elastic lines
  \cite{larkin_pinning_1979,robbins_contact_1987,demery_microstructural_2014,demery_effect_2014} allow us to link the hysteresis in apparent adhesion to the root-mean-square (rms) fluctuations of $w_\mathrm{loc}$.
  Inserting Eq.~\eqref{eq:KR-from-h-Fourier} into Eq.~\eqref{eq:G-perp-first-order}~and~\eqref{eq:W-equivalent2} yields
  \begin{equation}
  \label{eq:Wrms}
  w_\mathrm{rms} = \sqrt{\left\langle w_\mathrm{loc}^{2} - \left\langle w_\mathrm{loc} \right\rangle_{a, \theta}^{2} \right\rangle_{a,\theta}} = 2 \sqrt{e_\text{el} w_\text{int}}
  = h^{(1/2)}_\text{rms} \sqrt{E^{\prime} w_\text{int}}
  .
  \end{equation}
  The quantity $h_\text{rms}^{(1/2)}$ is the rms half-derivative (or quarter fractional Laplacian) given by
  \begin{equation}
      \left[h_\text{rms}^{(\alpha)}\right]^2
      =
      \frac{1}{4\pi^2}
      \int \dif^2 q \, |\vec{q}|^{2\alpha} C^{\mathrm{2D}}(\vec{q})
  \end{equation}
  with $\alpha=1/2$.

  The fluctuations of $w_\mathrm{loc}$ are linear in the roughness amplitudes.
  Since our assumption of fully conformal contact requires that $e_\text{el} < w_\text{int}$, $w_\mathrm{rms}$ is larger than the (second-order) shift in average adhesion.
  \M{link W to peaks}
  This strong \emph{linear} perturbation of the local energy arises because the
  solid is stretched by a distance
  $u(\xi) \propto  \sqrt{\xi} K/E^{\prime}$ with $K = \sqrt{2 w_\text{int}  E^{\prime}}$
  at an equilibrium crack tip.
  In valleys, the solid needs to stretch even more, increasing the elastic energy and decreasing the equivalent adhesion, while on roughness peaks, the equivalent adhesion \emph{increases} because the solid needs to stretch less than for a perfect sphere (see also Fig.~4 of the main text).
  The amplitude of these energy fluctuations are given by $h_\text{rms}^{(1/2)}$, a generalized measure of the sharpness of peaks sensitive to larger length scales than curvatures and slopes.
  For self-affine roughness, this parameter is dominated either by large scales like the rms height,
  or by small scales like slope and curvatures, depending on the Hurst exponent~\cite{persson_effect_2001}.

  
  \subsection{Non-circular contact: In-plane perturbation of the elastic energy release rate}
  \label{sec:in-plane}
  \begin{figure}
    \centering
     \includegraphics[width=\textwidth]{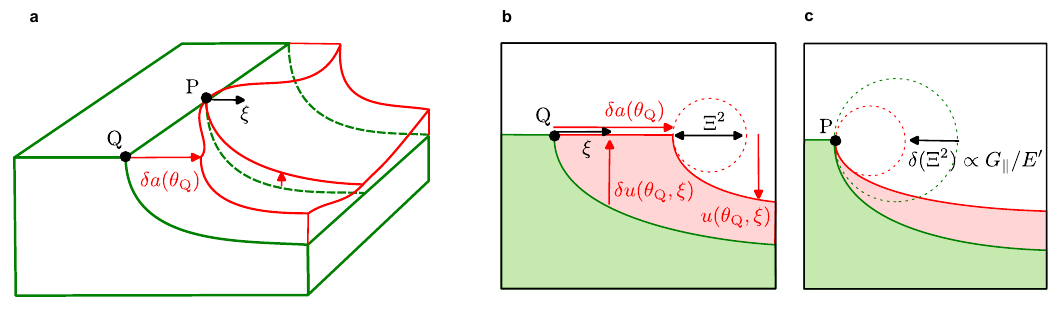}
     \caption{
       Effect of in-plane perturbations of the crack front
       on the energy release rate, $G_{\parallel}$.
       (a) The green solid represent a small section of the circular reference configuration with constant radius $a(\theta_{P})$ that we perturb by $\delta a(\theta_{Q}, \theta_{P}) = a(\theta_{Q}) -  a(\theta_{P}) $ (red).
       Advancing the contact area brings the crack faces closer together even in front of the point P that we hold fixed, because of the nonlocal interaction of the surface displacements.
       (b) At the crack tip, the displacements $u(\xi) \simeq \Xi \sqrt{\xi}$ with displacement intensity factor $\Xi$, so that
       closing the crack faces requires displacements $\delta u(\theta_{Q}, \xi) =  \Xi(\theta_{Q})\sqrt{\xi}$.
       The length $\Xi^{2}$ is the out-of-plane diameter at the crack tip (red circle) 
       and corresponds to the elastic energy release via $G\propto E^{\prime}\Xi^{2}$.
       At equilibrium, this diameter is proportional to the elastoadhesive length $\ell_{a}=w_\mathrm{int} / E^{\prime}$.
       (c)
       The diameter of the crack tip decreases by $\delta(\Xi^2) = 2 \Xi\Xi_{\parallel}$ as the crack faces come together at the point P.
     }
    \label{fig:in-plane-perturbation}
  \end{figure}
  Above, we discussed the effect of out-of-plane perturbation on a perfectly circular contact.
  In reality, the contact shape will deviate from circularity.
  We now compute
  the energy release rate at point P on a nearly circular contact to a rough sphere,
  \begin{equation}
      G([h,a]; b, \theta_\text{P})
      =
      G_\circ([h]; b, a(\theta_\text{P}), \theta_\text{P})
      +
      G_\parallel([a]; \theta_\text{P}),
  \end{equation}
  with first-order accuracy in the deviation from circularity $\delta a(\theta,\theta_\text{P}) = a(\theta) - a(\theta_\text{P})$, see Fig.~\ref{fig:in-plane-perturbation}.
  Our approximation is based on Gao and Rice's~\cite{gao_nearly_1987} first-order perturbation of the stress intensity factor at the perimeter of an initially circular external crack.
  %
  %
  Their result applies to arbitrary indenter geometries, where the stress intensity factor can vary along the perimeter~\cite{rice_three-dimensional_1985,borodachev_contact_1991,sanner_crack-front_2022},
  as is the case here due to surface roughness.
  We show that the first-order effect of the in-plane perturbation $G_{\parallel}$ is independent of the out-of-plane geometry,
  so that the contact of a rough sphere is equivalent to the contact of a smooth sphere discussed in Ref.~\cite{sanner_crack-front_2022} using the equivalent work of adhesion heterogeneity Eq.~\eqref{eq:W-equivalent2}.
  In Ref.~\cite{sanner_crack-front_2022}, we only considered the case where $G_\circ$ is uniform over the contact perimeter, such as when $G_\circ=G_\text{JKR}$.
  Here, we highlight the key changes required when $G_\circ$ is a function of $\theta$.
  
  The line elasticity emerges from the elastic coupling of the surface displacements caused by moving an initially stretched crack tip.
  The more the solid is stretched, the larger the elastic energy required to distort the contact line.
  In this section, we discuss perturbation of stress intensity factor in terms of displacements rather than pressures and introduce a displacement intensity factor~\cite{williams_stress_1957,hartranft_use_1969},
  \begin{equation}
  \label{eq:9}
  \Xi = \sqrt{8 / \pi} K / E^{\prime},
  \end{equation}
  in order to shorten the notation.
  Close to the crack tip, the geometry of the solid is described by
  \begin{equation}
  \label{eq:8}
  u(\xi, \theta) = \Xi(\theta) \sqrt{\xi} + \mathcal{O} (\xi^{3/2})
  ,
  \end{equation}
  so that
  \(\Xi^{2}\) corresponds to the diameter at the crack tip, see Fig.~\ref{fig:in-plane-perturbation}b.
  This length gives a geometric interpretation of the energy release rate via Eq.~\eqref{eq:Irwin},
  \begin{equation}
  \label{eq:Irwin-Xi}
  G = (\pi/16)  E^{\prime}  \Xi^{2}
  .
  \end{equation}
  At equilibrium, $G=w_\text{int}$, so that the diameter of the crack tip $\Xi_\mathrm{eq}^{2} = (16/\pi) \ell_\mathrm{a}$, with the elastoadhesive
  length $\ell_\mathrm{a} = w_\mathrm{int}/E^{\prime}$
  ~\cite{creton_fracture_2016}.
  
  Rice explained how distorting an initially circular contact perimeter affects the energy release rate in a point P that we hold fixed, see Fig.~\ref{fig:in-plane-perturbation}a.
  Making use of symmetries of the elastic potential,
  Rice showed that the crack-face weight-function describes how the surface of the solid moves as we distort the crack front within the plane. 
  %
  For the energy release rate, only the perturbation of crack-face displacements close to the crack tip matter.
  They are described by~\cite{gao_nearly_1987}
  \begin{equation}
  \label{eq:Xi-perturb}
  \begin{split}
  \Xi_{\parallel}([h,a]; b, \theta_{P})=
  &\Xi([h,a]; b, \theta_\text{P}) - \Xi_{\circ}([h]; b, a(\theta_\text{P}), \theta_\text{P})
  \\
  =&
  -
  \mathrm{PV} \int  \limits_{ 0}^{2\pi} \dif \theta_\text{Q}\,
    a(\theta_\text{P})  \frac{ (a(\theta_\mathrm{Q}) - a(\theta_\mathrm{P}) ) \Xi_\circ([h]; b, a(\theta_\text{Q}), \theta_\text{Q})}{||\vec{r_\text{P}} - \vec{r_\text{Q}}||^{2}}
    + \mathcal{O}([a - a(\theta_\text{P})]^{2}),
  \end{split}
  \end{equation}
  where $\Xi_{\circ}$ is the displacement intensity factor for the perfectly circular contact including roughness.
  The kernel of the integral was obtained from the $\xi \to 0$ limit of the crack-face weight-function of a circular external crack by Gao and Rice~\cite{gao_nearly_1987,wei_weight_1989,gladwell_contact_2008}.
  Equations~\eqref{eq:Irwin-Xi} and \eqref{eq:Xi-perturb} combined with the results from section~\ref{sec:effect-of-roughness} yield the energy release rate
  for the nearly circular contact to a rough sphere.

  Equation~\eqref{eq:Xi-perturb} captures the dominating effect of the long-ranged elastic coupling of the surface displacements on the energy release rate.
  We illustrate this effect in Fig.~\ref{fig:in-plane-perturbation}, where we hold $a(\theta_{P})$ fixed and advance the contact area in the neighborhood, corresponding to a locally convex perturbation of the contact perimeter (Fig.~\ref{fig:in-plane-perturbation}a).
  Closing the adhesive neck
  over the surface element
  $\dif\theta\, a(\theta_\mathrm{Q})  \delta a(\theta_\mathrm{Q})$
  requires the displacement $\delta u (\theta_\mathrm{Q}, \xi) = \Xi(\theta_\mathrm{Q}) \sqrt{\xi}$ (Fig.~\ref{fig:in-plane-perturbation}b),
  which perturbs the whole surface of the solid with amplitudes decaying with distance as $||\vec{r}_{P} - \vec{r}_{Q}||^{-2}$.
  We hold the crack front locally pinned in $\theta_{P}$,
  yet this nonlocal interaction along the crack front brings the crack faces together (Fig.~\ref{fig:in-plane-perturbation}c)
  and thereby reduces the energy release rate $G(\theta_{P})$.
  
  We now show that within our first-order approximation, the reduction of $G$ for convex $a$ discussed in the previous paragraph
  is independent of the indenter geometry and stiffness, and
  corresponds to a generalized curvature,
  the half-fractional Laplacian  $( - \Delta_{s} )^{1/2} a(s)$, where $s=a \, \theta$ is an arclength along the contact perimeter.
  The principal value integral in Eq.~\eqref{eq:Xi-perturb} depends on roughness, indenter geometry and indenter position via $\Xi_{\circ}$, but this coupling of the in-plane elastic response to the out-of-plane geometry is only $\mathcal{O}(\delta a^{2})$.
  Near equilibrium, where
  \begin{equation}
  \label{eq:Xi0-near-equilibrium}
  \Xi_{\circ}([h]; b, a(\theta), \theta) + \mathcal{O}([\aeq-\aeq(\theta)]) =
  \Xi_\mathrm{eq}
  = \sqrt{\frac{16 w_\mathrm{int}}{\pi E^{\prime}}}
  ,
  \end{equation}
  deviations of $\Xi_{\circ}$ from the material property $\Xi_\mathrm{eq}$
  manifest in Eq.~\eqref{eq:Xi-perturb} through the second-order term $\Xi_\circ \delta a$.
  Approximating $\Xi_{\circ}$ by the constant value $\Xi_\mathrm{eq}$ simplifies the principal value integral to
  \begin{equation}
  \label{eq:Xi-pure-first-order}
  \begin{split}
  \Xi_{\parallel}([a];\theta_\text{P})
  \approx
  - \Xi_\mathrm{eq} \,
  \mathrm{PV} \int  \limits_{ 0}^{2\pi} \dif \theta_\text{Q}\,
  a(\theta_\text{P})  \frac{\delta a(\theta_\text{Q})}{||\vec{r_\text{P}} - \vec{r_\text{Q}}||^{2}}
  = \Xi_\mathrm{eq}  \left( - \Delta_{s} \right)^{1/2} a(\theta_{\mathrm{P}}).
  \end{split}
  \end{equation}
  Here, the half-fractional Laplacian of the contact radius with respect to the arclength $\dif s= a(\theta_\text{P}) \dif \theta_\text{P}$ is defined by
  \begin{equation}
    \left( - \Delta_{s} \right)^{1/2}a(\theta_{\mathrm{P}})
   = \frac{1}{a(\theta_{\mathrm{P}})}\left( - \Delta_{\theta} \right)^{1/2} a(\theta_{\mathrm{P}})
  = \frac{1}{a(\theta_{\mathrm{P}})}
    \sum \limits_{n \in \mathbb{Z}^{\backslash \{0\}}}
    |n| \ \tilde a_n \ e^{i n \theta_{\mathrm{P}}},
  \end{equation}
  where $\tilde{a}_{n}$ are the coefficients of the Fourier series
  \begin{equation}
  \label{eq:5}
  a(\theta) = \sum_{\mathrm{Z}} \tilde a_n \ e^{i n \theta}.
  \end{equation}
  The wavelength of a Fourier mode is $\ell_{n} =  2\pi a(\theta) / |n|$.
  The Fourier amplitude of $\left( - \Delta_{s} \right)^{1/2}a(\theta_{\mathrm{P}})$, $\tilde a_{n} / \ell_{n}$, is the slope of the Fourier mode, but unlike slopes, the maxima and minima of the fractional Laplacian are in phase with maxima and minima of $a$.
  Hence, $\left( - \Delta_{s} \right)^{1/2}a$ can be interpreted as a generalized curvature scaling like a slope.
  
  Equations~\eqref{eq:Xi0-near-equilibrium} and \eqref{eq:Xi-pure-first-order} yield the first-order perturbation of the energy release rate
  \begin{equation}\label{eq:G_first_order}
  \begin{split}
    G_{\parallel} ([a], \theta_\mathrm{P})
    \approx
    \frac{\pi}{8} \Xi_\mathrm{eq} \Xi_{\parallel}([a], \theta_\mathrm{P})
    =
    w_{\mathrm{int}} \left( - \Delta_{s} \right)^{1/2} a( \theta_{\mathrm{P}}),
  \end{split}
  \end{equation}
  %
  %
  describing that the line penalizes deviations from circularity with a strength proportional to the equilibrium energy release rate $w_\text{int}$ and a generalized curvature.
  This means that for a fixed jump-depth $\delta a = d$,
  it is easier to deflect the line over a wider lateral section $\ell$, $\delta G = w_\text{int} d /  \ell$,
  explaining why a row of several asperities can \emph{collectively} pin the crack front while an individual asperity cannot~\cite{imry_random-field_1975}.
  
  \section{Numerical implementation of the crack-front model}
  \label{sec:numerical-cf}
  
  Our numerical simulations use the algorithm by Rosso and Krauth~\cite{rosso_roughness_2002} to solve for the equilibrium configurations (metastable states) visited by the crack front as we pull the sphere in and out of the contact. 
  We discretize the crack front in $N$ collocation points at equally spaced angles $\theta$ following Ref.~\cite{sanner_crack-front_2022}.
  
  The surface roughness $h(x,y)$ is a Gaussian random field, 
  where the height spectrum $\tilde{h}(q_x, q_y)$ has uncorrelated phases 
  and random amplitudes scaling according to the PSD, 
  and defines the equivalent work of adhesion field via Eqs.~\eqref{eq:KR-from-h-Fourier} and ~\eqref{eq:W-equivalent2}.
  Equation~\eqref{eq:KR-from-h-Fourier} describes the stress intensity factor for a straight crack that is rotated to be tangential to the contact circle. 
  Note that the prefactor in Eq.~\eqref{eq:KR-from-h-Fourier} is a complex number that 
  introduces a minor phase-shift between $w_\mathrm{loc}$ and $h$ in
  the direction normal to the front.
  While this phase-shift, and thereby the orientation of the crack,
  are important when comparing deterministically the crack-front model to the BEM, 
  they have no effect on the power-spectrum of $w_\mathrm{loc}$
  and on the work of adhesion hysteresis.
  %
  When the correlation length is much smaller than the contact radius, 
  the heights decorrelate before the orientation of the crack changes significantly along the perimeter. 
  For this reason, and because the rotation becomes computationally intractable on large grids,
  we generate the equivalent work of adhesion fields used in the main text and in Suppl. Sec.~\ref{sec:parametric-study} using a constant orientation of the crack.

  \section{Validation of the crack-front model against the Boundary Element Method}
  \label{sec:cf-bem}
  
  \begin{figure*}
    \centering
    \includegraphics{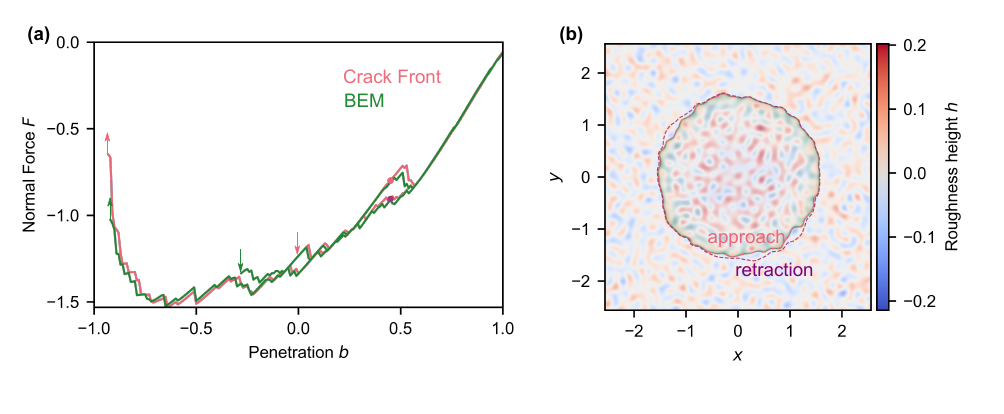}
    \caption{
    \textbf{(a)} Force-penetration curves from a boundary element method (BEM) and a crack-front simulation on the random roughness shown in panel (b).
      The arrows indicate the jump into contact and the jump out of contact instabilities.
      \textbf{(b)} Contact areas at the penetration $b=0.45$, indicated by the dots in the force-penetration curve,
      on top of the surface topography.
      The tensile pressures of the contact mechanics simulation during approach are shown in green,
      so that the perimeter of the contact is indicated by the darkest green pixels.
    The dashed lines represent the contact perimeter calculated with the crack-front model during approach (pink) and retraction (purple).
   The BEM simulation was discretized on a $1024 \times 1024$ grid with pixel size $\ell_\mathrm{pix} = 0.005$.
  The roughness is a random Gaussian field with a flat power spectrum at wavelengths above the cutoff wavelength $\lambda_\mathrm{r}=0.2$ and 0 below. 
  The interaction is a cubic polynomial with a cutoff distance $g_c=0.24$, corresponding to a cohesive zone size $\ell_\mathrm{coz} = (\pi/36) g_c^2 / \ell_a \simeq 0.012$.
  In both simulations, we increased the penetration $b$ in steps of 0.01
  until the maximum penetration $b_\mathrm{max} = 1$ was reached
  and then decreased it until pull off.
  %
  The results are nondimensionalized following the conventions of Refs.~\protect\cite{barthel_adhesive_2008,maugis_contact_2010}.
    }
    %
      \label{fig:cf-bem}
  \end{figure*}
   
  We compare the crack-front model to a boundary element method (BEM) simulation to validate our mapping from surface roughness to an equivalent work of adhesion heterogeneity.
  The implementation of the BEM and the parameters of the simulation are similar to Ref.~\cite{sanner_crack-front_2022}, 
  where we validated the crack-front model for spheres with heterogeneous work of adhesion.
  In the BEM simulation we perform here, the sphere is rough and the work of adhesion is uniform.
  The surfaces interact with a cubic cohesive law with a hard-wall repulsion.
  Our implementation of the BEM is described in detail in Ref.~\cite{sanner_crack-front_2022} and is based on \cite{hockney_potential_1970,byrd_limited_1995,stanley_fft-based_1997,campana_practical_2006,pastewka_contact_2016}.
  %

  
  Figure~\ref{fig:cf-bem} shows a BEM and a crack front simulation on 
  random roughness with $e_\mathrm{el}/w_\mathrm{int} \simeq 0.03$
  and a power spectrum that is flat for wavelengths above the correlation length $\lambda_\mathrm{r}=0.2$
  and 0 below. 
  The force-penetration curves computed with the BEM and the crack-front model nearly overlap and contact perimeters agree well, confirming that the contact of rough spheres is equivalent to the pinning of a crack by the work of adhesion heterogeneity $w_\mathrm{loc}$ mapped using equation~\eqref{eq:W-equivalent2}.
  %
  Note that in the BEM, the jump into contact instability occurs too early because of the finite interaction range~\cite{wu_jump--contact_2010, ciavarella_effect_2017, wang_modeling_2021,sanner_crack-front_2022}. 
  This particular event converges slowly with interaction range, while the remainder of the force-penetration curve,
  including depinning instabilities, is well converged.
  %
  %
  %
  %
  Other discrepancies in the force-penetration curves are due to the linearization in the crack-front model.

  \section{Hysteresis on random roughness: crack-front simulations}
  \label{sec:parametric-study}

  We verify our theoretical prediction for the apparent work of adhesion (Main Text Eq.~(6))
  \begin{equation}
  \label{eq:hysteresis_roughness_suppl}
  w_\mathrm{\overset{\scriptstyle retr}{ \scriptstyle appr}}
  =
  w_\mathrm{int} - e_\mathrm{el} \pm k e_\mathrm{el}
  ,
  \end{equation}
  using crack-front simulations on self-affine roughness with varying power spectra,
  and extract the numerical factor $k\approx 3$ from these results.
  We show in Suppl. Sec. I B,C that self-affine surface roughness maps to an equivalent work of adhesion field with power-law correlation via the integral transform Eq.~\eqref{eq:KR-from-h-Fourier}.
  The variance of this work of adhesion heterogeneity $w_\mathrm{rms}^{2} = 4 e_\mathrm{el} w_\mathrm{int}$, with $e_\mathrm{el}$ the elastic energy required to conform to the surface roughness.
  
\begin{figure}
    \centering
    \includegraphics[]{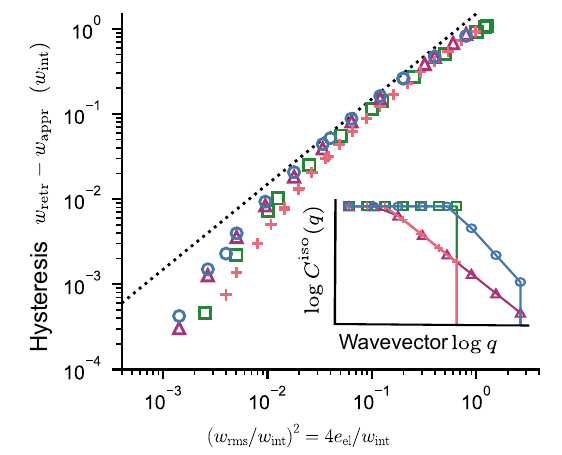}
    \caption{
      Crack front simulations on random fields of adhesion $W(x,y)$ show that the hysteresis in apparent work of adhesion $w_\mathrm{retr} - w_\mathrm{appr} \propto w_\mathrm{rms}^{2}$ (dashed line).
      %
      %
      We show in Suppl. Sec. I B,C that surface roughness with elastic energy for fully conformal contact $e_\mathrm{el}$ is equivalent to a work of adhesion heterogeneity with variance $w_\mathrm{rms}^{2} = 4 e_\mathrm{el} w_\mathrm{int}$.
      The work of adhesion fields used here correspond to randomly rough surfaces with different shapes of the power spectra represented in the inset.
      %
      %
      %
      We used a flat PSD with short-wavelength cutoff $\lambda_{s}=0.005$ (green squares),
      and three different self-affine PSDs
      parameterized by the rolloff wavelength $\lambda_{r}$, the short-wavelength cutoff $\lambda_{s}$ and the Hurst exponent $H$.
      The self-affine PSDs scale with $ |q|^{-2-2H}$ for wavelengths between $\lambda_{s}$ and $\lambda_{r}$.
      The blue circles correspond to $\lambda_{s} = 0.000625, \lambda_{r} = 0.01, H=0.8$; the purple triangles to $\lambda_{s} = 0.000625, \lambda_{r} = 0.1  , H=0.3$;
      and the pink crosses to $\lambda_{s} = 0.005, \lambda_{r} = 0.1 , H=0.3$.
      The force-area curve for the blue circle at $e_\mathrm{el}/w_\mathrm{int} = 0.05$ is shown in Main Text Fig.~3c.
    }
    \label{fig:parametric-study}
  \end{figure}

  Figure~\ref{fig:parametric-study} shows the work of adhesion hysteresis $w_\mathrm{retr} - w_\mathrm{appr}$
  as a function the amplitude of disorder $(w_\mathrm{rms}/w_\mathrm{int})^{2} = 4 e_\mathrm{el} / w_\mathrm{int}$.
  The dashed line represent the prediction Eq.~\eqref{eq:hysteresis_roughness_suppl} using $k\approx 3$ that we fitted to the results.
  For $e_\mathrm{el}/w_\mathrm{int} \gtrsim 0.01 $, the work of adhesion hysteresis in our numerical simulations (symbols) overlaps with the theoretically predicted scaling (dashed line)
  and is independent of the shape of the power spectrum.
  Below a critical value of $e_\mathrm{el}$, the hysteresis disapears because of the finite size of the contact~\cite{robbins_contact_1987,tanguy_weak_2004,demery_microstructural_2014}.
  This onset of hysteresis depends on the shape of the power-spectral density:
  for roughness with a short correlation length (purple triangles),
  the contact process starts to dissipate energy at smaller $e_\mathrm{el}$ than for a longer correlation length (pink crosses).
  
  The scaling of the hysteresis with $w_\mathrm{rms}^{2}$ was theoretically predicted and numerically verified on random fields with short-ranged correlation~\cite{larkin_pinning_1979,demery_microstructural_2014,demery_effect_2014},
  similar to our roughness
  with flat PSD represented by the green squares.
  Here we consider isotropic self-affine roughness leading to work of adhesion fields with long-ranged power-law correlations.
  D\'emery et al.~\cite{demery_effect_2014} theoretically predicted that for isotropic fields, the relationship between hysteresis and $w_\mathrm{rms}$ remains unaffected by these power-law correlations.
  They derived this result by analytically solving a small disorder expansion of the equation of motion of the elastic line.
  Our numerical simulations further confirm that Eq.~\eqref{eq:hysteresis_roughness_suppl} remains valid for isotropic self-affine roughness.
  
  \begin{figure}
     \centering
     \includegraphics[scale=1.42]{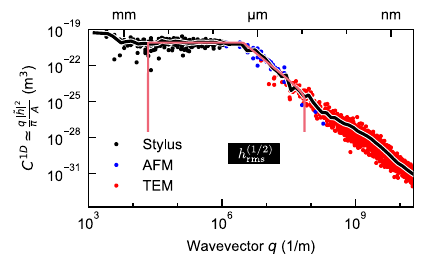}
     \caption{
       One dimensional power-spectral density of the nanocrystalline rough surface and of the synthetic surface used in the simulation shown in main-text Fig.~5c.
       We fit an ideal self-affine PSD with hurst exponent $H=1$ to the experimental data (light red line).
       The rolloff wavelength is $2.1\,\mu$m and we chose the amplitude of the PSD to match the elastic energy for fully conformal contact, giving the amplitude of the 2D PSD in the rolloff region is $8.4\cdot 10^{-27}\,$m.
       The vertical lines show the length scales included in our simulation.
     }
     \label{fig:sm_fig6}
  \end{figure}
  

\end{document}